# Система обработки и анализ трекинг- данных для исследования пути клиента на основе RFID-технологии

**М.В. Холод**

E-mail: Kholod.MV@rea.ru

ФГБОУ ВО «Российский экономический университет им. Г.В. Плеханова», Москва, Россия

**Аннотация**

Статья посвящена исследованию системы обработки и анализа трекинговых данных на основе RFID-технологии для изучения пути клиента (customer journey) в розничной торговле. Рассматривается эволюция RFID-технологии, её ключевые принципы работы и современные применения в ритейле, выходящие за рамки логистики и включающие точное управление запасами, борьбу с потерями и улучшение клиентского опыта. Особое внимание уделяется архитектуре сбора, обработки и интеграции данных, в частности, методологии ETL (extract, transform, load) для преобразования сырых RFID и POS-данных в структурированное аналитическое хранилище. Предлагается детализированная логическая модель базы данных, предназначенная для комплексного анализа, объединяющего финансовые метрики продаж с поведенческими паттернами перемещения покупателей. В статье также анализируются ожидаемые бизнес-преимущества от внедрения RFID через призму системы сбалансированных показателей (BSC), оценивающей финансовую эффективность, клиентскую удовлетворенность и оптимизацию внутренних процессов. Делается вывод о том, что интеграция трекинговых и транзакционных данных создает основу для превращения ритейла в точную, основанную на данных науку, обеспечивая беспрецедентную видимость физических потоков товаров и поведения потребителей.

**Ключевые слова:** данные треков отслеживания клиентов, RFID-технология, цифровая платформа управления, ритейл-аналитика, путь клиента, ETL, хранилище данных, система сбалансированных показателей (BSC).

**Введение**

Внедрение радиочастотной идентификации (RFID – radio frequency identification) в ритейле, начавшееся в 2000-х годах, было сопряжено со значительными трудностями. Многие амбициозные проекты, такие как тотальное размечивание всего товара в сетях, например, Metro Group, не достигли первоначальных целей из-за высоких затрат на метки, технической незрелости стандартов и проблем с интеграцией в сложные IT-ландшафты. Однако современная цифровая трансформация бизнеса, в том числе и ритейла, создала принципиально новые предпосылки для ренессанса технологии. Повсеместный переход на электронный документооборот, использование больших данных, онлайн-касс, искусственного интеллекта и чат-ботов сформировало среду, где точные и автоматизированные данные RFID стали не просто опцией, а критически важным активом для принятия решений в реальном времени. Современная розничная торговля переживает трансформацию по причине цифровизации и интеграции продвинутых трекинговых технологий. Среди них радиочастотная идентификация (RFID) занимает ключевое место, обеспечивая возможность сбора данных о поведении потребителей и движении товаров в режиме реального времени. Данный обзор систематизирует существующие исследования по применению RFID-технологий в розничной торговле и фокусируется на анализе поведения потребителей, управлении запасами и внедрении технологий. RFID-технология представляет собой форму автоматической идентификации и сбора данных (AIDC), которая использует электромагнитные поля на радиочастотах для однозначной идентификации объектов, их аутентификации, точного определения местоположения и автоматической передачи данных. Сферы ее применения в торговле простираются от управления цепочками поставок и активов до инвентаризации в режиме реального времени и полностью автоматизированных платежных решений (как, например, платные дороги или системы доступа). В части применения технологии RFID в современном ритейле помимо базового отслеживания паллет и коробок, современные RFID-системы решают более сложные задачи, такие как точная инвентаризация (accuracy ~98%) в реальном времени. Ритейлеры проводят инвентаризацию всего торгового зала и склада за считанные часы, а не недели, просто пронеся считыватель RFID меток вдоль стеллажей. Это решает проблему «несогласованности данных» по остаткам или по ценам в информационной системе и на физическом товаре. В свою очередь это позволяет снизить уровень отсутствия товарных позиций на полках или складе до 80%, что напрямую конвертируется в увеличение продаж, прибыль и повышение доли рынка.

Также, RFID-системы позволяют вести борьбу с потерями товаров и метки на каждом товаре позволяют отслеживать подозрительные перемещения и идентифицировать источники краж как со стороны посетителей, так и персонала. Системы могут активировать аварийную сигнализацию при выносе немаркированного товара или отправлять уведомления о несанкционированном перемещении товара.

Кроме того, RFID-системы помогают улучшать клиентский опыт через умные примерочные, оснащенные RFID-считывателями, автоматически распознают вынесенные товары и выводят на экран информацию о размерах, цветах, доступных альтернативах и стилях. Это увеличивает средний чек и вовлеченность покупателя.

Автоматизация логистики на складе реализована через RFID-ворота, которые на приемке автоматически сверяют поступивший товар с заказом-накладной, значительно ускоряя процесс и минимизируя человеческий фактор. Далее система точно знает, где находится каждая единица товара, оптимизируя процессы комплектации и пополнения полок.

Техническим ядром системы является RFID-метка — микрочип с антенной, которая крепится на тару (паллеты, коробки) или непосредственно на товар. Метка хранит уникальный электронный код продукта (EPC), который считывается дистанционно специальным ридером. Ридер преобразует радиосигнал в цифровые данные, которые затем интегрируются в корпоративные информационные системы (WMS, ERP) для дальнейшей обработки и анализа. Способность технологии достаточно точно и автоматически идентифицировать каждый объект предоставляет компаниям беспрецедентные возможности для оптимизации. Архитектура RFID-системы концептуально состоит из двух взаимосвязанных подсистем - подсистемы, выполняющей идентификацию и транзакции с использованием беспроводной связи и включающую в себя метки и считыватели, которые могут быть стационарные, мобильные и встроенные в оборудование; а также корпоративной (предприятийная) подсистемы, включающей серверы, промежуточное ПО (Middleware), базы данных и аналитические системы, в задачи которой входит фильтрация, агрегация, хранение и преобразование сырых RFID-данных в пригодную для бизнес-анализа информацию, интегрированную в такие процессы, как управление запасами, CRM и аналитика [8]. Также может существовать организационная подсистема, которая обеспечивает обмен информацией между различными участниками цепочки создания стоимости (производителями, дистрибьюторами, логистическими операторами, ритейлерами) через стандартизированные протоколы такие как информационные сервисы электронного кода продукта (EPCIS). Это позволяет отслеживать движение маркированного товара на всем его жизненном цикле — от производства до конечной продажи и даже после нее (например, для программ утилизации или возврата).

Таким образом, современная RFID-система — это не просто набор считывателей и меток, сложная экосистема, которая обеспечивает цифровую видимость всего физического потока товаров, превращая ритейл в точную науку на основе данных.

## 1. Теоретические и практические основы использования трекинг-данных клиентов из RFID-систем в ритейле

RFID технологии прошли длинный путь своего развития с момента их изобретения. Как отмечают [1], технология прошла путь от военных применений до массового использования в логистике и розничной торговле. [2] и [3] детализируют технические принципы работы RFID систем, включая классификацию меток (активные/пассивные), частотные диапазоны и протоколы связи. В современных работах, таких как [4], особое внимание уделяется развитию безчиповым RFID технологиям, которые обещают дальнейшее снижение стоимости и расширение областей применения. В работе [5] предоставляется практическое руководство по внедрению RFID, делая технологию доступной для технических специалистов и инженеров. В соответствии с таблицей 1 ключевым применением RFID в розничной торговле является трекинг потребительского поведения. [6] и [7] демонстрируют, как RFID данные позволяют анализировать движение покупателей по торговому залу, время остановок в отделах, и частоту посещения отделов. Эти данные, интегрированные с POS-информацией, создают целостное представление о пути клиента, как отмечено в работе [8] и [9]. Также в работе [10] идентифицируются драйверы и барьеры персонализации на основе RFID данных, подчеркивая важность интеграции данных и их конфиденциальность. Кроме того, RFID технологии позволяют перейти от традиционных тепловых карт к интеллектуальному управлению пространством. В работе [11] описывается, как данные о движении клиентов позволяют оптимизировать выкладку товаров, планировку магазина и размещение рекламных материалов. [12] показывает, что RFID системы не только улучшают клиентский опыт, но и способствуют снижению потерь и повышению операционной эффективности.

Многочисленные исследования подтверждают влияние RFID на эффективность управления цепями поставок. Работы [13] и [14] отмечают улучшение видимости цепей поставок, снижение ошибок и ускорение процессов. Также в работах [15] и [16] демонстрируется, что внедрение RFID приводит к значительному улучшению точности управления запасами, снижая затраты на складские операции. В промышленном контексте [17] и [18] показывают эффективность RFID для управления запасами в сталелитейной промышленности. Также современные тенденции включают интеграцию RFID с IoT платформами. [19] и [20] описывают, как RFID датчики становятся частью интеллектуальных сред, позволяющие осуществлять мониторинг в режиме реального времени и управление. [21] рассматривают современные методы RFID позиционирования, которые повышают точность трекинга объектов и людей. [22] анализируют диффузию RFID технологий в индустрию моды, отмечая возрастающий уровень адаптации данной технологии к этой индустрии и появление новых приложений на основе технологии.

Таблица 1 Обзор исследований применения трекинговых (RFID) технологий в различных сферах бизнеса

| Сфера использования RFID-технологий | Ожидаемая ценность для бизнеса | Исследования |
|---|---|---|
| Логистика и управление цепочками поставок | Автоматическая идентификация и отслеживание грузов, паллет, контейнеров на всём пути от производителя до склада. Снижение логистических издержек до 15%, повышение прозрачности цепи поставок, исключение ошибок ручного ввода, сокращение времени обработки грузов. | Sarac, A., Absi, N., & Dauzère-Pérès, S. (2010) Zelbst, P. J., et al. (2012) |
| Розничная торговля и управление запасами | Точный учёт товаров в режиме реального времени на складах и в торговых залах. Автоматизация приёмки, инвентаризации и предотвращения краж. Повышение точности учёта запасов до ~99%, сокращение затрат на проведение инвентаризации на 80%, снижение уровня (отсутствия товара (out-of-stock) до 50%. | De Marco, A., et al. (2012) Gaukler, G. M., & Seifert, R.W. (2007) |
| Управление активами и производством | Отслеживание инструментов, оборудования, незавершённого производства на заводе. Контроль доступа к | Zhou, S., & Ling, W. (2013) Nayak, R., et al. (2015) |

| | критическим активам. Повышение эффективности использования активов, сокращение времени на поиск оборудования, предотвращение потерь и краж дорогостоящего инструмента. | |

Внедрение инновационных технологий, таких как RFID, в ритейле представляет собой сложный и капиталоемкий процесс, требующий тщательного обоснования и последующего контроля. На рис. 1 показана система сбалансированных показателей внедрения трекинговых технологий (на примере RFID-технологий) как инструмент оценки их эффективности. Традиционные финансовые метрики, хотя и остаются критически важными, не в полной мере отражают стратегические выгоды от подобных проектов, такие как повышение клиентской лояльности или оптимизация операционных процессов. В этой связи система сбалансированных показателей предлагает целостную концепцию, позволяющую связать технологические инвестиции со стратегическими целями компании. Ключевой целью внедрения RFID с финансовой точки зрения является обеспечение окупаемости инвестиций (ROI) и повышение общей прибыльности бизнеса. Эмпирические данные свидетельствуют о том, что крупные розничные сети могут достигать ежегодной экономии в миллиарды долларов за счет комплексного эффекта от технологии. Ключевыми индикаторами успеха в данной перспективе выступают: показатель ROI, целевое значение которого может варьироваться в диапазоне 20-30% в течение двух лет; сокращение потерь от краж на уровне до 75%; снижение операционных затрат на логистику и проведение инвентаризаций на 30%; а также увеличение объема продаж на 5-10% благодаря повышению уровня доступности товара на полках. С точки зрения клиентской перспективы стратегической целью в данной области является трансформация клиентского опыта и усиление лояльности. RFID-технологии способствуют достижению этой цели за счет значительного ускорения процессов обслуживания, в частности, сокращения времени на кассовые операции на 50%, а также за счет повышения точности выполнения онлайн-заказов (click-and-collect) до 98%. В результате, компании фиксируют рост удовлетворенности клиентов, который может быть количественно измерен через специализированные опросы (NPS или CSI) с целевым показателем улучшения на 20%. Это создает долгосрочное конкурентное преимущество, выражающееся в увеличении клиентской базы и частоты покупок.

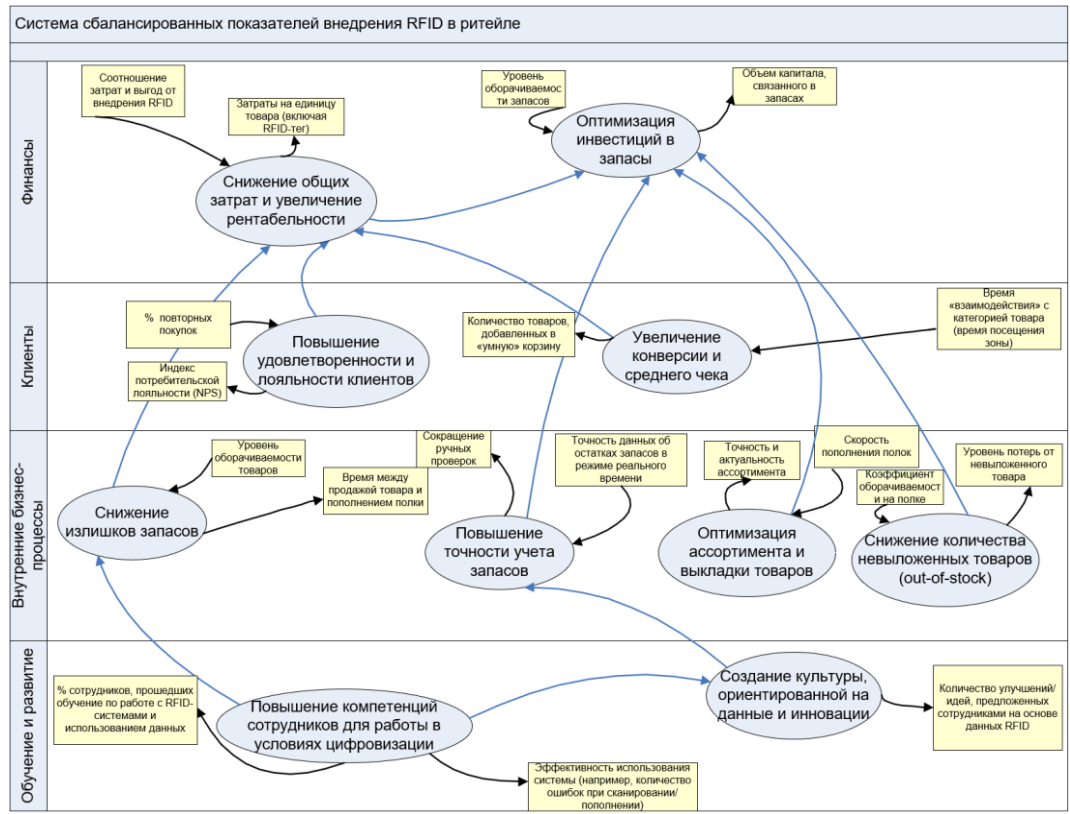

Рис. 1. Система сбалансированных показателей внедрения трекинговых технологий (на примере RFID-технологий) как инструмент оценки их эффективности

В разрезе внутренних бизнес-процессов, повышенная видимость цепочек поставок и поведения покупателей обеспечивает возможность оптимизации ассортимента товаров и эффективного пополнения полок. Снижение необходимой полочной площади позволяет сократить объем запасов на складах в любой момент времени, что высвобождает оборотный капитал и создает условия для диверсификации ассортимента.

## 2. Архитектура сбора данных треков отслеживания клиентов в ритейле

Для аналитики данных была использована общая схема (*рис.2*), изображающая традиционную архитектуру бизнес-аналитики и хранилища данных. На изображении показан поток данных от источника, через процесс очистки и преобразования, и, наконец, в структурированный формат, оптимизированный для анализа и составления отчетов, дашбордов и моделей данных.

Архитектура следует классическому шаблону ETL («извлечение, преобразование, загрузка») для перемещения данных из информационных систем CRM или ERP в аналитическую базу данных (хранилище данных) и, наконец, в многомерную модель куба данных для бизнес-пользователей.

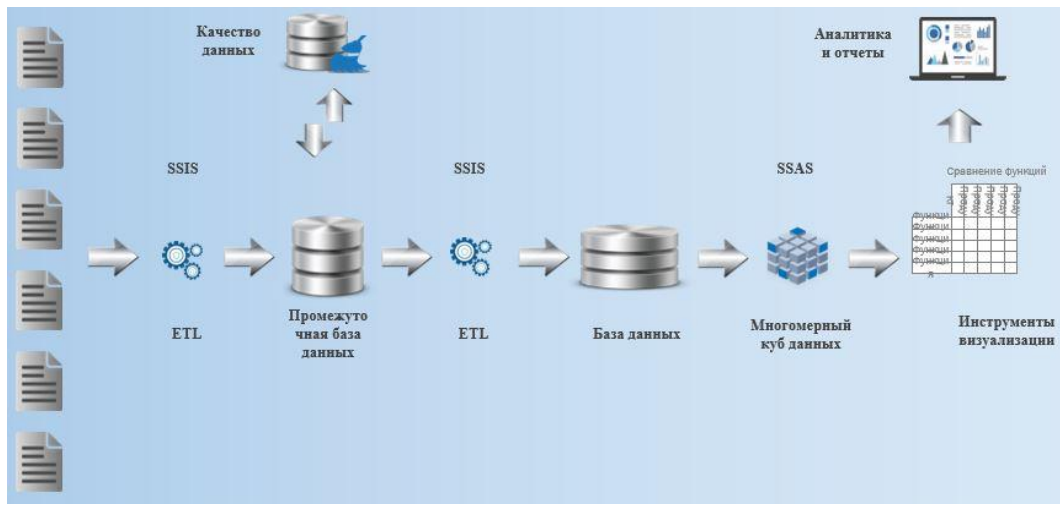

Рис. 2. Техническая архитектура сбора, хранения и обработки данных треков отслеживания клиентов

При этом основной задачей при создании подобных архитектур является обеспечение качества данных. Это отправная точка и ключевой принцип, реализацию которого необходимо обеспечить при создании технической архитектуры трекинг-данных клиентов. Он представляет собой процесс профилирования, очистки, проверки и обогащения необработанных данных, поступающих из исходных систем, перед их поступлением в основной конвейер ETL, который является автоматизированным процессом перемещения из различных источников в целевую систему, и состоящий из 3 этапов – извлечения этих данных, преобразования их в нужный формат и последующей их загрузки для анализа. Такие конвейеры обеспечивают качество данных, позволяя компаниям принимать обоснованные решения на основе консолидированных, организованных и полных данных. В целях практической реализации это часто достигается с помощью таких инструментов, как сервисы обеспечения качества данных (DDS). В данной работе используется SSIS, одна из основных ETL-систем, которая позволяет реализовать весь процесс перемещения и преобразования данных. При этом данные проходят этап ETL дважды. На первом этапе (первый SSIS) выполняется начальное извлечение из исходных систем, первый этап очистки данных, их разметка и преобразование данных на основе правил качества данных. На втором этапе (второй SSIS) берутся проверенные данные из промежуточной базы данных и на основании бизнес-логики совершаются преобразования данных, например, объединение строк RFID-данных, в которых зафиксированы посекундные перемещения клиентов по плоскости магазина, и другие преобразования, а также создание вычисляемых столбцов, в которых, например, агрегировано время перемещения в разрезе одного отдела супермаркета. После этого, данные загружаются в окончательное структурированное хранилище данных. Промежуточная база данных представляет собой временное пространство для загрузки сырых данных, извлечённых из исходных систем. Структура промежуточной базы данных обычно отражает исходные системные таблицы. Основные задачи промежуточной базы данных состоят в том, чтобы изолировать процесс извлечения от процесса преобразования, обеспечить быструю загрузку данных без сложных преобразований, служить точкой восстановления в случае сбоя процесса ETL. Основная база данных создана для аналитических целей и представляет собой структурированную размеченную систему данных, состоящую из таблиц фактов. Например, в случае с RFID-данными каждая строка в основной базе данных соответствует факту перемещения клиента в пространстве магазина с автоматизированным замером его x-,y-координат на плоскости магазина, а также времени, рассчитанного в промежуточной базе данных на основе автоматически зафиксированных отметок времени, проведенного клиентом в данной точке. Кроме того, в основной базе данных содержатся такие описательные атрибуты из базы сырых транзакционных POS-данных как наименование товара, который клиент купил в данном отделе в данный момент времени. Эти данные подтянуты в систему из источника сырых POS-данных. Следующим этапом является формирование многомерного куба (куба данных), с помощью которого считываются структурированные данные из хранилища данных, также происходит агрегирование и расчет метрик. Это семантический уровень, обеспечивающий удобное для бизнеса представление и анализ данных одновременно с нескольких точек зрения (измерений). Затем используются инструменты для создания отчётов, визуализации и моделирования данных для получения результатов конечных пользователей. При этом системы визуализации и моделирования напрямую подключаются к кубу или к хранилищу данных SSAS для создания интерактивных панелей мониторинга, диаграмм и отчётов, расчетов метрик или модельных показателей. Это последний уровень, предоставляющий аналитические данные бизнес-аналитикам, менеджерам и лицам, принимающим решения.

Логическая модель хранилища данных представлена на рис.3. Она является центральным компонентом в общей архитектуре, так как именно ее структура определяет, какие виды анализа и отчеты смогут построить бизнес-пользователи

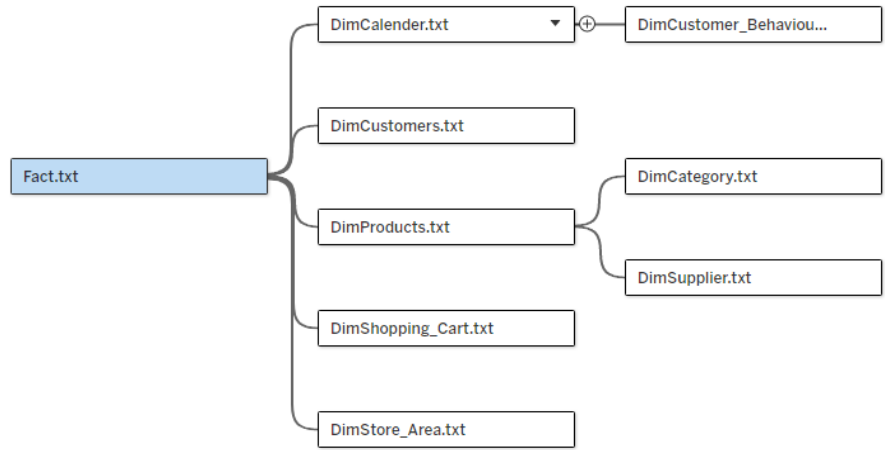

Рис. 3. Логическая модель основного хранилища RFID и POS-данных

. Текстовые файлы содержат иерархии для анализа по дате, которые связаны с поведенческими статусами клиентов, а именно со статусом передвижения клиентов по плоскости магазина – sus (остановки) и mig (миграции), а также меток времени, в которое данные передвижения происходили. Далее используется файл, содержащий демографические и статистические данные по клиентам, включая ID клиента. Далее – файл, который описывает атрибуты корзины покупок с данными по товарной иерархии и компаний-поставщиков. Далее данные по координатам x- и y- перемещения корзин покупок по плоскости магазина, а также отдельный файл с последовательностью секций магазина в соответствии с его картой.

3. **Описание аналитической базы данных розничного магазина на основе POS и RFID-данных**

Для комплексного анализа бизнес-метрик необходима продуманная основа, объединяющая традиционные финансовые метрики (полученные на основе транзакционных данных ID-POS) с передовыми данными о поведении клиентов в физическом пространстве (полученных на основе RFID-данных), что особенно актуально для задач розничной аналитики (retail analytics) и анализа цепочек взаимодействия (customer journey analysis).

На рисунке 4 представлена детализированная логическая модель витрины данных, спроектированная для поддержки аналитических процессов в предметной области ритейл-аналитики, связанной с розничной торговлей и анализом поведения потребителей.

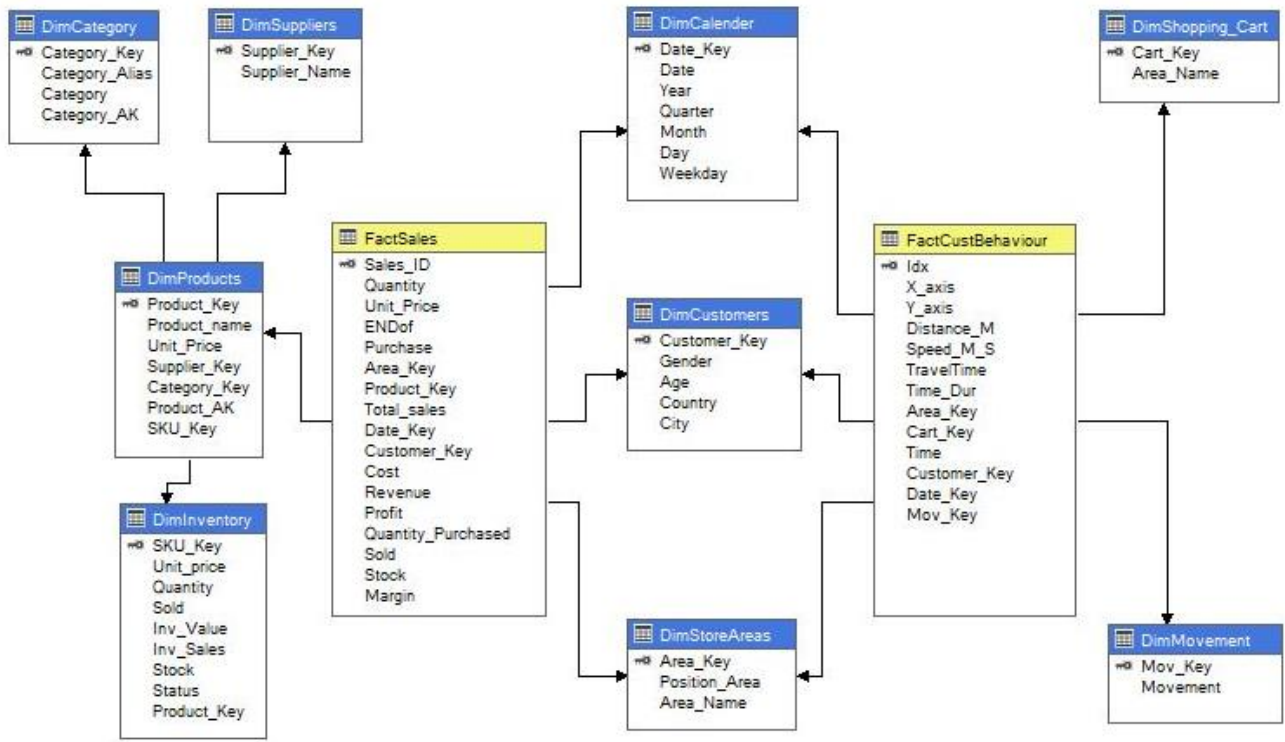

Рис. 4. Логическая модель структуры данных для поддержки аналитики на основе POS и RFID-данных

Структура аналитической базы данных состоит из набора таблиц-измерений (Dm — Dimension) и таблиц-фактов (Facts), которые хранят бизнес-метрики. Данная организация обеспечивает целостность данных, минимизирует избыточность и оптимизирует выполнение сложных запросов с агрегацией по множеству измерений.

Центральным элементом схемы выступает фактная таблица FactSales, аккумулирующая ключевые количественные показатели (меры) бизнес-деятельности. В нее входят метрики объема операций (Quantity, Quantity_Purchased, Sold), финансовые показатели (Cost, Revenue, Profit, Total_sales, Margin), а также операционные данные о состоянии запасов (Stock). Для связи с контекстуальными измерениями таблица содержит набор внешних ключей (Product_Key, Customer_Key, Date_Key, Area_Key).

Окружающие фактную таблицу измерения предоставляют аналитический контекст. Измерение DmProducts описывает ассортимент товаров, включая атрибуты наименования, цены и артикула (SKU_Key), и связано с измерением поставщиков DmSupplies. Стандартизированное измерение DmCalender с иерархией Дата > День > Месяц > Квартал > Год обеспечивает возможности для временного анализа. Особый интерес представляет структура данных о клиентах. Модель разделяет эту сущность на две логические компоненты: классическое демографическое измерение DmCustomers (пол, возраст, географическое местоположение) и специализированную фактную таблицу, обозначенную как FactCustBehavior, которая содержит динамические данные о пространственно-временной активности внутри помещений. Последняя включает координаты (X-axis, Y-axis), скорость перемещения (Speed_M_S), пройденное расстояние (Distance_M) и временные метки, полученные на основании технологий RFID-трекинга.

Для интерпретации данных введены вспомогательные измерения, такие как Area_Key, включающие наименования зон магазин (Area_Name), а измерение DmMovement классифицирует тип перемещения (Movement). Наличие связи между активностью клиента, зонами и фактами продаж создает мощную основу для анализа эффективности мерчандайзинга и поведения потребителей. предложенная модель данных поддерживает многомерный анализ по широкому спектру направлений: от традиционного анализа продаж и рентабельности по товарам, поставщикам и времени до продвинутого поведенческого анализа, связывающего паттерны перемещения клиентов с их покупательским поведением.

В процессе реализации аналитической модели данных, необходимо интегрировать независимые источники транзакционных данных (POS-данных) с трекинговыми данными (RFID-данными) для формирования целостного профиля поведения потребителя. На рисунке 5 показана концептуальная модель интеграции двух независимых источников данных, что является основой для углубленного поведенческого анализа. Модель демонстрирует переход от изолированного изучения финальной покупки к комплексному исследованию всего пути клиента (customer journey) внутри торговой точки.

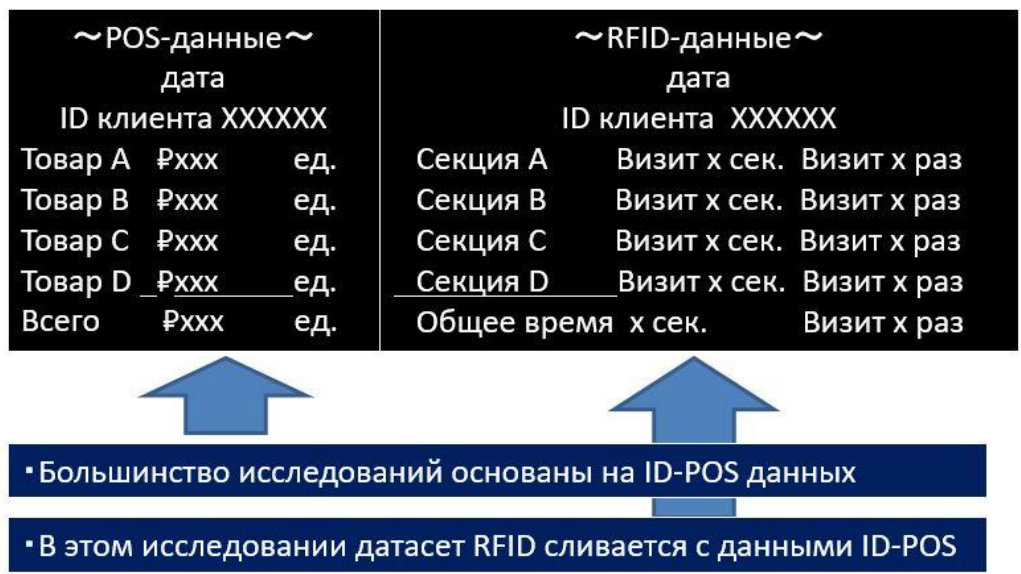

Рис. 5. Схематическая визуализация методологии обогащения (через ID клиента) стандартных транзакционных данных (POS-данных) трекинговыми данными (RFID-данными) для формирования целостного профиля клиентского поведения

Левая часть рисунка иллюстрирует традиционный источник данных — транзакционные данные (POS-данные). Данная структура представляет собой детализированный чек, агрегированный по уникальному идентификатору клиента (Клиент IDXXXXXX) и временной метке (год, месяц, день). Содержательная часть таблицы включает перечень приобретенных товаров (Товар A, Товар B, и др.); финансовые метрики, такие как стоимость каждой позиции и общая сумма чека; количественные метрики, такие как количество приобретенных единиц по каждой позиции и итоговое количество товаров.

Эти данные отвечают на вопросы «что?», «сколько?» и «по какой цене?» клиент купил в конкретный момент времени, но не дают информации о процессе принятия решения.

Правая часть рисунка представляет более современный источник данных — трекинговые данные (RFID-данные), собранные с помощью радиочастотной идентификации. Эти данные структурированы для того же клиента (Клиент IDXXXXXX) и за тот же временной период (год, месяц, день), что обеспечивает возможность их интеграции. При этом таблица трекинговых данных содержит следующие поведенческие метрики: метку локация, а именно идентификатор отдела или зоны магазина (отдел A, отдел B и пр.); временные метрики, а именно длительность остановки (Остановка . сек.) и общее время движения (Общее время движения x сек.), а также метрики посещаемости, такие как количество визитов в конкретную зону (посещение x раз) и общее количество посещений всех зон.

## Заключение

Проведенное исследование демонстрирует, что внедрение RFID-технологий кардинально трансформирует современный ритейл, выводя его на качественно новый уровень управления на основе данных. Из инструмента для автоматической идентификации RFID эволюционировал в комплексную экосистему, обеспечивающую сквозную цифровую видимость всего жизненного цикла товара и поведения потребителя.

Ключевым результатом работы является разработка архитектуры и логической модели интеграции разнородных данных. Предложенный подход, основанный на ETL-процессах и организации хранилища данных, позволяет объединить традиционные транзакционные данные (POS) с инновационными трекинговыми данными (RFID). Это создает целостную основу для анализа пути клиента (customer journey), связывая паттерны перемещения по торговому залу с финальными покупками. Такая интеграция дает ритейлерам беспрецедентную возможность понять не только *что* купил клиент, но и *как* он принимал это решение, какие зоны магазина привлекли его внимание и повлияли на выбор.

Анализ эффективности внедрения через систему сбалансированных показателей (BSC) подтверждает значительный стратегический эффект от технологии. RFID вносит вклад не только в операционную эффективность (сокращение потерь от краж на 75%, снижение затрат на логистику и инвентаризацию на 30%, точность учета запасов до 99%), но и в финансовые результаты (рост продаж на 5-10% за счет снижения уровня out-of-stock) и, что наиболее важно, в повышение лояльности и удовлетворенности клиентов.

Таким образом, (AI) для прогнозной аналитики, что откроет новые возможности для персонализации реализация предложенной системы обработки и анализа трекинг-данных на основе RFID переводит розничную торговлю из области эмпирических решений в область точной, data-driven науки. Дальнейшее развитие связано с интеграцией RFID с технологиями Интернета Вещей (IoT) и искусственного интеллекта клиентского опыта и полной автоматизации управления розничным предприятием.

## Литература

**Об авторах**

**Холод Марина Викторовна**

Кандидат экономических наук, доцент РЭУ им. Г.В. Плеханова;

ведущий научный сотрудник, Дирекция по науке и инновациям РЭУ им. Г.В. Плеханова, Россия, 115054, г. Москва, Стремянный переулок, д.36

E-mail:
Kholod.MV@rea.ru

ORCID: 0000-0002-4738
3534

# The system of processing and analysis of customer tracking data for customer journey research on the base of RFID technology

**M.V. Kholod**

E-mail: Kholod.MV@rea.ru

Plekhanov Russian University of Economics, Moscow, Russia

**Abstract**

The article focuses on researching a system for processing and analyzing tracking data based on RFID technology to study the customer journey in retail. It examines the evolution of RFID technology, its key operating principles, and modern applications in retail that extend beyond logistics to include precise inventory management, loss prevention, and customer experience improvement. Particular attention is paid to the architecture for data collection, processing, and integration, specifically the ETL (extract, transform, load) methodology for transforming raw RFID and POS data into a structured analytical data warehouse. A detailed logical database model is proposed, designed for comprehensive analysis that combines financial sales metrics with behavioral patterns of customer movement. The article also analyzes the expected business benefits of RFID implementation through the lens of the Balanced Scorecard (BSC), which evaluates financial performance, customer satisfaction, and internal process optimization. It is concluded that the integration of tracking and transactional data creates a foundation for transforming retail into a precise, data-driven science, providing unprecedented visibility into physical product flows and consumer behavior.

**Keywords:** customer tracking data, RFID technology, digital management platform, retail analytics, customer journey, ETL, data warehouse, Balanced Scorecard (BSC)

**Citation:** Kholod, M.V. (2025). A System for Processing and Analyzing Tracking Data to Research the Customer Journey Based on RFID Technology. Business Informatics, Vol. ___, No. ___, pp. _____.

This research was performed in the framework of the state task in the field of scientific activity of the Ministry of Science and Higher Education of the Russian Federation, project "Models, methods, and algorithms of artificial intelligence in the problems of economics for the analysis and style transfer of multidimensional datasets, time series forecasting, and recommendation systems design", grant no. FSSW-2023-0004.

**Об авторах**

Холод Марина Викторовна

Кандидат экономических наук, доцент РЭУ им. Г.В. Плеханова;

ведущий научный сотрудник, Дирекция по науке и инновациям РЭУ им. Г.В. Плеханова,

Россия, 115054, г. Москва, Стремянный переулок, д.36

E-mail: Kholod.MV@rea.ru

ORCID: 0000-0002-4738 3534